\documentclass[10pt]{iopart}
\usepackage{graphicx}
\usepackage{latexsym}
\usepackage{url}

\usepackage{iopams}
\def\aap{A\&A\,  }
\def\apj{ApJ\,  }
\def\apjl{ApJ\,  }

\def\mnras{MNRAS\,  }

\def\sun{\hbox{$\odot$}}
\begin{document} 
\title
{
Standard and Truncated Luminosity Functions 
for  stars in the Gaia Era
}
\vskip  1cm
\author     {Lorenzo  Zaninetti}
\address    {Physics Department,
 via P.Giuria 1,\\ I-10125 Turin,Italy }
\ead {zaninetti@ph.unito.it}

\begin {abstract}
The  luminosity function (LF) for stars  is here  fitted
by a Schechter  function and by a Gamma probability
density function.
The dependence of the number of stars
on the distance, both in the low and high luminosity
regions, requires the  inclusion of a lower and upper
boundary in  the Schechter and Gamma LFs.
Three astrophysical applications for stars  are provided:
deduction of the parameters at low distances,
behavior  of the average absolute magnitude
with distance,  and the location
of  the photometric maximum
as a function of the selected flux.
The use of  the truncated LFs allows to model
the  Malmquist bias.
\end{abstract}
\vspace{2pc}

\maketitle

\section{Introduction}

The stellar luminosity function (LF) 
is  the relative numbers
of stars  of different luminosities in a
standard volume of space ,usually a cubic
parsec.
The determination of the LF
for  stars is complicated at a  local  level
by the presence of five classes
for the stars, as given by the MK system,
and by the mass-luminosity relation.
The presence of the Malmquist bias,
after \cite{Malmquist_1920,Malmquist_1922,Malmquist_1936},
for an introduction, see section 3.6
in \cite{Binney1998}
or the historical section 2  in \cite{Butkevich2005},
modifies  the  distribution in absolute magnitude
as a function of the distance and therefore
complicates  the modeling  of the LF for stars.

The  LFs  for stars started to be  fitted
by a Gaussian  probability density function (PDF)
in  absolute  magnitude,
see \cite{Eddington1914}.
In order to deal with the boundaries,
a  double
truncated Gaussian  in absolute magnitude
has  been  considered, see \cite{Jaschek1985}.
The astronomical  derivation of the LF 
takes account of   
a standard volume
with a radius of $\approx 20\,pc$.
As an example \cite{Wielen1974} has derived the first local 
LF for stars in a spherical volume having radius of $22\,pc$
and more recently \cite{Flynn2006}
has  measured the volume luminosity density and surface luminosity density
generated by the Galactic disc, using accurate data on the local luminosity
function and the vertical structure of the disc. 
A new sample of stars, representative of the solar neighborhood LF, 
has been constructed  from the  Hipparcos (HIP)  catalogue and the Fifth
Catalogue of Nearby Stars, 
see \cite{Just2015}.

From the previous  analysis, the
following questions can be  raised.
\begin{itemize}
\item
Is it possible to
model the LF for stars with the Schechter function
and the Gamma LF?
\item
Is it possible to model the absolute magnitude-distance
plane with the truncated  Schechter function
or the truncated Gamma LF?
\item
Is it possible to model
the  observational maximum in the number of stars
and the average number of stars
versus
distance at a given flux?
\end{itemize}

\section{The Gaia Catalog}

A great number of stars
with mean apparent magnitude in
the G-band, flux, $f$,
expressed in electron-charge per second (e-/s)
and parallax,
$\approx$ two million,
are available
at the  Gaia Data Release 1 (Gaia DR1)
astrometric catalogs, see   \cite{GAIA2016a,GAIA2016b},
with data  at
\url{http://vizier.u-strasbg.fr/viz-bin/VizieR}
and  specific Table I/337/tgasptyc.
The above catalog gives  
stellar parallax,
G-band  flux,
G-band magnitude,
Tycho-2 or HIP     BT magnitude
and Tycho-2 or HIP VT magnitude.
As pointed out    by    \cite{Stassun2016}  
there is  an average offset of $-0.25 \pm 0.05$ mas in  the Gaia
parallaxes   and therefore we increased by 0.25 the parallax.
According to Gaia DR1,
the luminosity as deduced from the flux
will be expressed in Gaia units, namely, 
$e-/s\,pc^2$.
The $G$ magnitude, see \cite{GAIA2017b}
 is
\begin{equation}
G = -2.5 \log (f) + zp
\quad ,
\end{equation}
where $zp$ is the photometric zero
derived as  in \cite{GAIA2016c}, we found numerically
$zp=25.52$.

\begin{figure}
\begin{center}
\includegraphics[width=7cm]{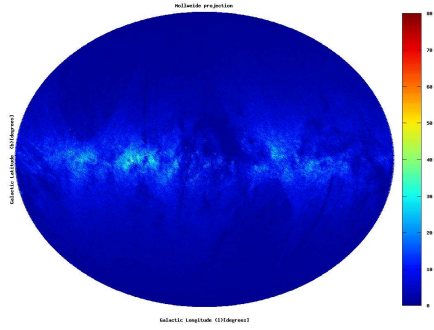}
\end{center}
\caption
{
Mollweide projection  of the sky density of
all Gaia DR1 sources in Galactic coordinates.
}
 \label{gaia_mollweide}%
\end{figure}

The distribution of all Gaia DR1 sources in the sky is illustrated
in Figure \ref{gaia_mollweide}.

\begin{figure}
\begin{center}
\includegraphics[width=7cm]{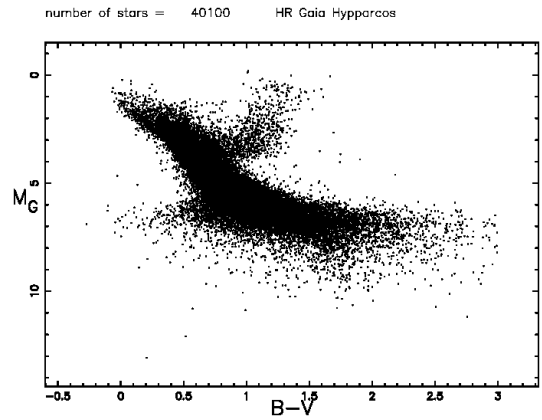}
\end{center}
\caption
{
$M_{\mathrm G}$  against $(B-V)$
, evaluated as  BT-VT,
(H-R diagram)
in the first  100~pc.
}
 \label{gaia_hr}%
\end{figure}

\begin{figure}
\begin{center}
\includegraphics[width=7cm]{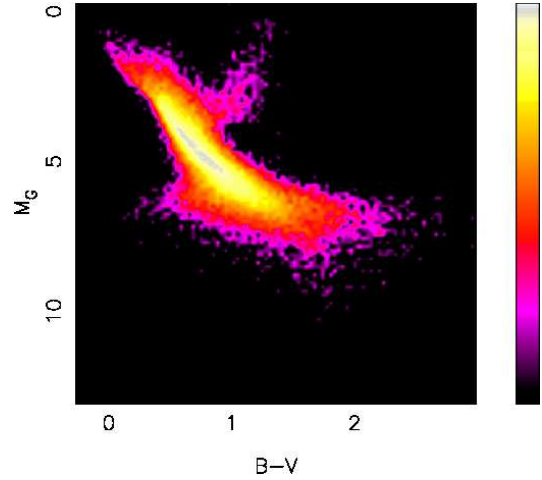}
\end{center}
\caption
{
Contour density of stars for  H-R diagram
in a logarithmic  scale.
}
 \label{gaia_hr_contour}%
\end{figure}

The observational Hertzsprung-Russell
(H-R) diagram
in $M_G$ as obtained by the Gaia DR1
parallaxes
versus (B-V) , evaluated as  BT-VT, 
is  presented in
Figure \ref{gaia_hr}
and  in a contour density version
in
Figure \ref{gaia_hr_contour},
see also
figure 1 in \cite{GAIA2017a}.

\begin{figure}
\begin{center}
\includegraphics[width=7cm]{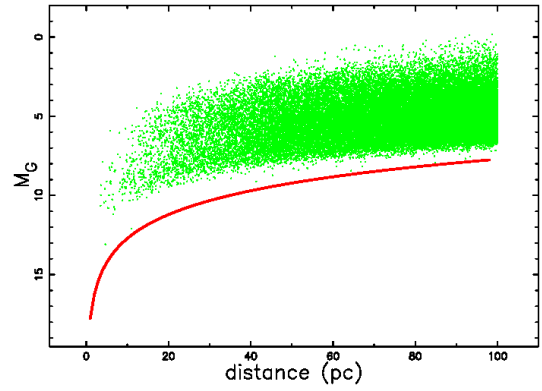}
\end{center}
\caption
{
$M_{\mathrm G}$  versus  distance in pc
in the first  100~pc             (green points)
and  theoretical   upper  curve in magnitude
lower  curve in the plot (red  line)
when $m_G$=12.71.
}
 \label{gaia_lower}%
\end{figure}

The distance  modulus is
\begin{equation}
m_G - M_G =  5\,\log (d) -5
\quad  ,
\label{distancemodulusstars}
\end{equation}
where
$m_g$ is the  apparent magnitude in  the G-band,
$M_g$ is the  absolute magnitude in  the G-band
and  $d$ is the distance  in pc.
Isolating  $M_G$ in the above equation
we obtain  the theoretical curve for the
upper observable absolute  magnitude
\begin{equation}
M_g =-5\,\log (d) +5  +m_G
\quad ,
\label{mabsgupper}
\end{equation}
once the maximum  apparent magnitude 
in the g-band, $m_{lim}$, 
is inserted,
i.e. $m_G$=12.71.
Figure   \ref{gaia_lower}  presents
the absolute magnitude  as function of the
distance as well the  upper theoretical curve
in magnitude.

\begin{figure}
\begin{center}
\includegraphics[width=7cm]{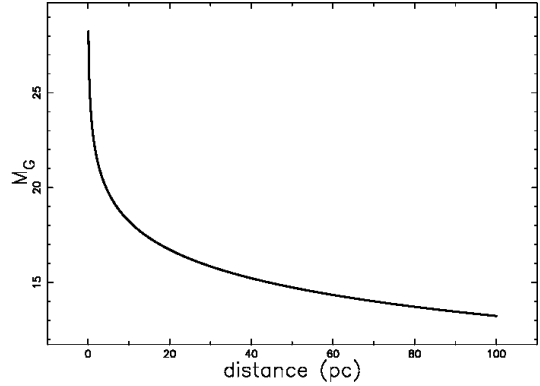}
\end{center}
\caption
{
The  relationship of completeness for 
$M_{\mathrm G}$  versus  distance in pc.
}
 \label{sample_complete}%
\end{figure}

The completeness of the sample can be evaluated by the 
following relationship for the absolute magnitude 
\begin{equation}
M_g= -{\frac {-{\it m_{lim}}\,\ln  \left( 10 \right) +5\,\ln  \left( d
 \right) -5\,\ln  \left( 10 \right) }{\ln  \left( 10 \right) }}
\quad  .
\end{equation}
On inserting  in the above formula 
$m_{lim}$=12.71 we  obtain a numerical relationship
between selected absolute magnitude 
and numerical relationship over which
the sample  is complete, see Figure \ref{sample_complete}.

In the case here considered 
the absolute  magnitude 
covers the range   $[3 , 12 \, mag ]$
and therefore we deal with a complete sample.

\section{Standard LFs}

Here we introduce  
an algorithm to build the LF,
the statistical tests adopted,
as well as the Schechter and Gamma LFs.
The derived parameter for the local LF 
will be  applied in Section \ref{secaverage}
according to the general principle
that the LF is equal everywhere 
but the upper observable absolute magnitude 
decreases with distance.

\subsection{The astronomical LF}

A LF for stars  is  built  
according to the following points
\begin{enumerate}
\item 
A standard  distance is  chosen, i.e. $20\,pc$,
\item 
The GAIA's stars  are selected according
to the following  ranges of existence:
$-5 \leq M_V \leq 15$ where $M_V$  is the absolute visual 
magnitude  and  $-0.3 \leq (B-V) \leq 3$,
\item  We organize an histogram
with bins large 1 mag 
\item  
We divide the  obtained frequencies by the 
involved volume,
\item
We do not  apply  the  $1/V_{a}$ method because
our sample is complete at $20\,pc$, 
\item 
The error  of the LF is evaluated as the square root
of the frequencies divided by the  involved volume. 
\end{enumerate}

\begin{figure}
\begin{center}
\includegraphics[width=7cm]{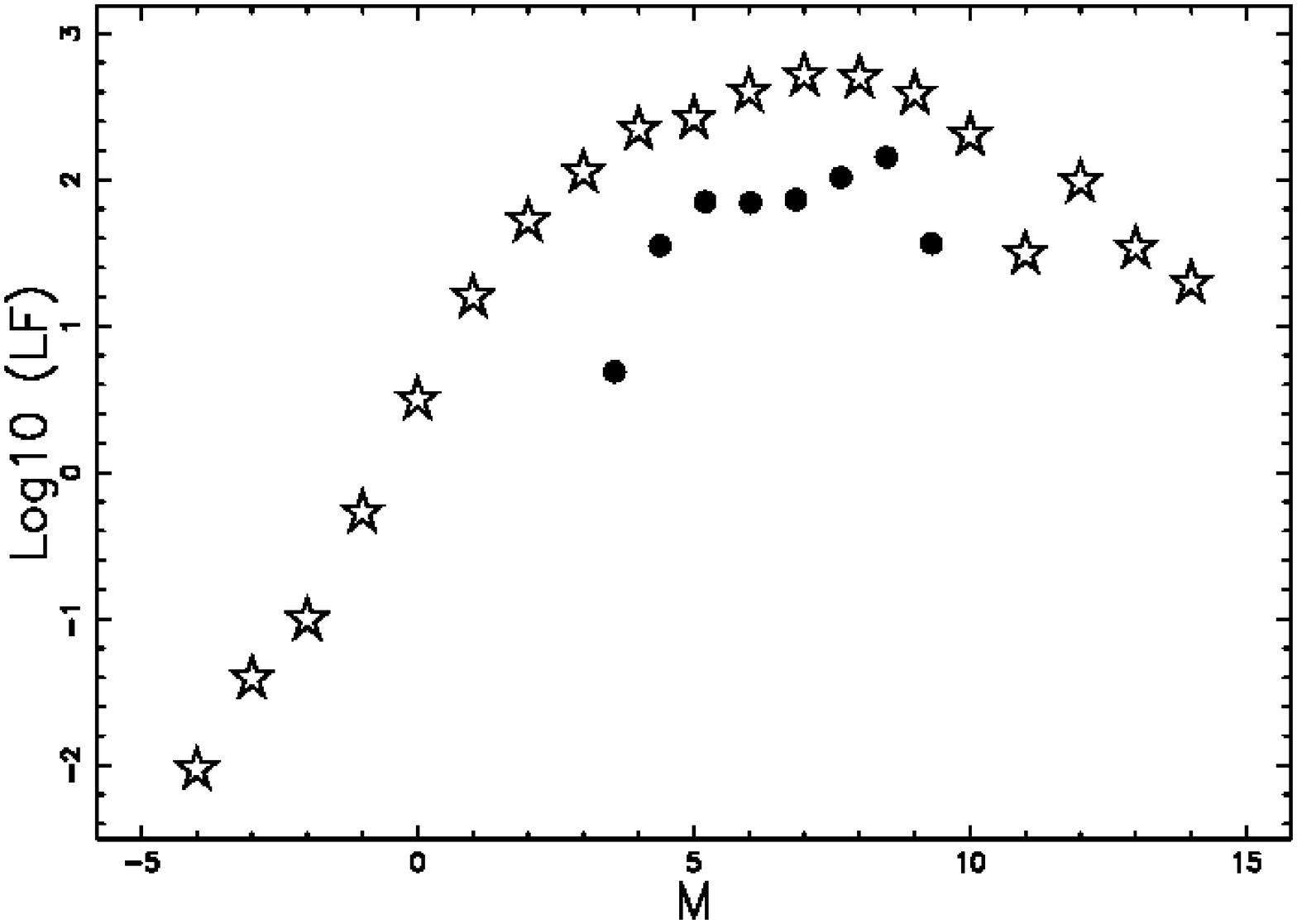}
\end{center}
\caption
{
LF in the V band main sequence, 
empty stars,  
and Gaia's LF, filled circles.
}
 \label{gaia_lf_due}%
\end{figure}

The LF for Gaia's  stars  is
reported in Figure \ref{gaia_lf_due}
together the LF main sequence in the V band 
as extracted from 
Table 2, column 9, in \cite{Just2015}.

\subsection{Statistical Tests}

The merit function $\chi^2$
is  computed as
\begin{equation}
\chi^2 =
\sum_{j=1}^n ( \frac {LF_{theo} - LF_{astr} } {\sigma_{LF_{astr}}})^2
\quad ,
\label{chisquare}
\end{equation}
where   $n$ is the number of bins for the LF of the stars  and the two
indices $theo$ and $astr$ stand for `theoretical'
and `astronomical', respectively.
The reduced  merit function $\chi_{red}^2$
is  evaluated  by
\begin{equation}
\chi_{red}^2 = \chi^2/NF
\quad,
\label{chisquarereduced}
\end{equation}
where $NF=n-k$ is the number of degrees  of freedom
and $k$ is the number of parameters.
The goodness  of the fit can be expressed by
the probability $Q$, see  equation 15.2.12  in \cite{press},
which involves the number of degrees of freedom
and $\chi^2$.
According to  \cite{press}, the
fit ``may be acceptable'' if  $Q \geq 0.001$.
The Akaike information criterion
(AIC), see \cite{Akaike1974},
is defined by
\begin{equation}
AIC  = 2k - 2  ln(L)
\quad,
\end {equation}
where $L$ is
the likelihood  function  and $k$ is  the number of  free parameters
in the model.
We assume  a Gaussian distribution for  the errors
and  the likelihood  function
can be derived  from the $\chi^2$ statistic
$L \propto \exp (- \frac{\chi^2}{2} ) $
where  $\chi^2$ has been computed by
Equation~(\ref{chisquare}),
see~\cite{Liddle2004}, \cite{Godlowski2005}.
Now the AIC becomes
\begin{equation}
AIC  = 2k + \chi^2
\quad.
\label{AIC}
\end {equation}

\subsection{The Schechter LF}

Let $L$, the  luminosity of a star,
be defined in $[0, \infty]$.
The Schechter LF of the stars, $\Phi$, originally applied to the stars,
see \cite{schechter},
is
\begin{equation}
\Phi (L;\Phi^*,\alpha,L^*) dL =
(\frac {\Phi^*}{L^*}) (\frac {L}{L^*})^{\alpha}
\exp \bigl ( {- \frac {L}{L^*}} \bigr ) dL \quad,
\label{lf_schechter}
\end {equation}
where $\alpha$ sets the slope for low values
of $L$,
$L^*$ is the
characteristic luminosity, and $\Phi^*$ represents
the number of stars per pc$^3$.
The  normalization is
\begin{equation}
\int_0^{\infty} \Phi (L;\Phi^*,\alpha,L^*) dL  =
\rm \Phi^*\, \Gamma \left( \alpha+1 \right)
\quad  ,
\label{norma_schechter}
\end{equation}
where
\begin{equation}
\rm \Gamma \, (z )
=\int_{0}^{\infty}e^{{-t}}t^{{z-1}}dt
\quad ,
\end{equation}
is the Gamma function.
The average luminosity,
$ { \langle L \rangle } $, is
\begin{equation}
{ \langle (\Phi (L;\Phi^*,\alpha,L^*) \rangle }
=
\rm L^* \,{\rm \Phi^*  }\,\Gamma \left( \alpha+2 \right)
\quad  .
\label{ave_schechter}
\end{equation}
An equivalent form  in absolute magnitude
of the Schechter LF
is
\begin{eqnarray}
\Phi (M;\Phi^*,\alpha,M^*)dM=
\nonumber\\
0.921 \Phi^* 10^{0.4(\alpha +1 ) (M^*-M)}
\exp \bigl ({- 10^{0.4(M^*-M)}} \bigr)  dM \, ,
\label{lfstandard}
\end {eqnarray}
where $M^*$ is the characteristic magnitude.

\begin{figure}
\begin{center}
\includegraphics[width=7cm]{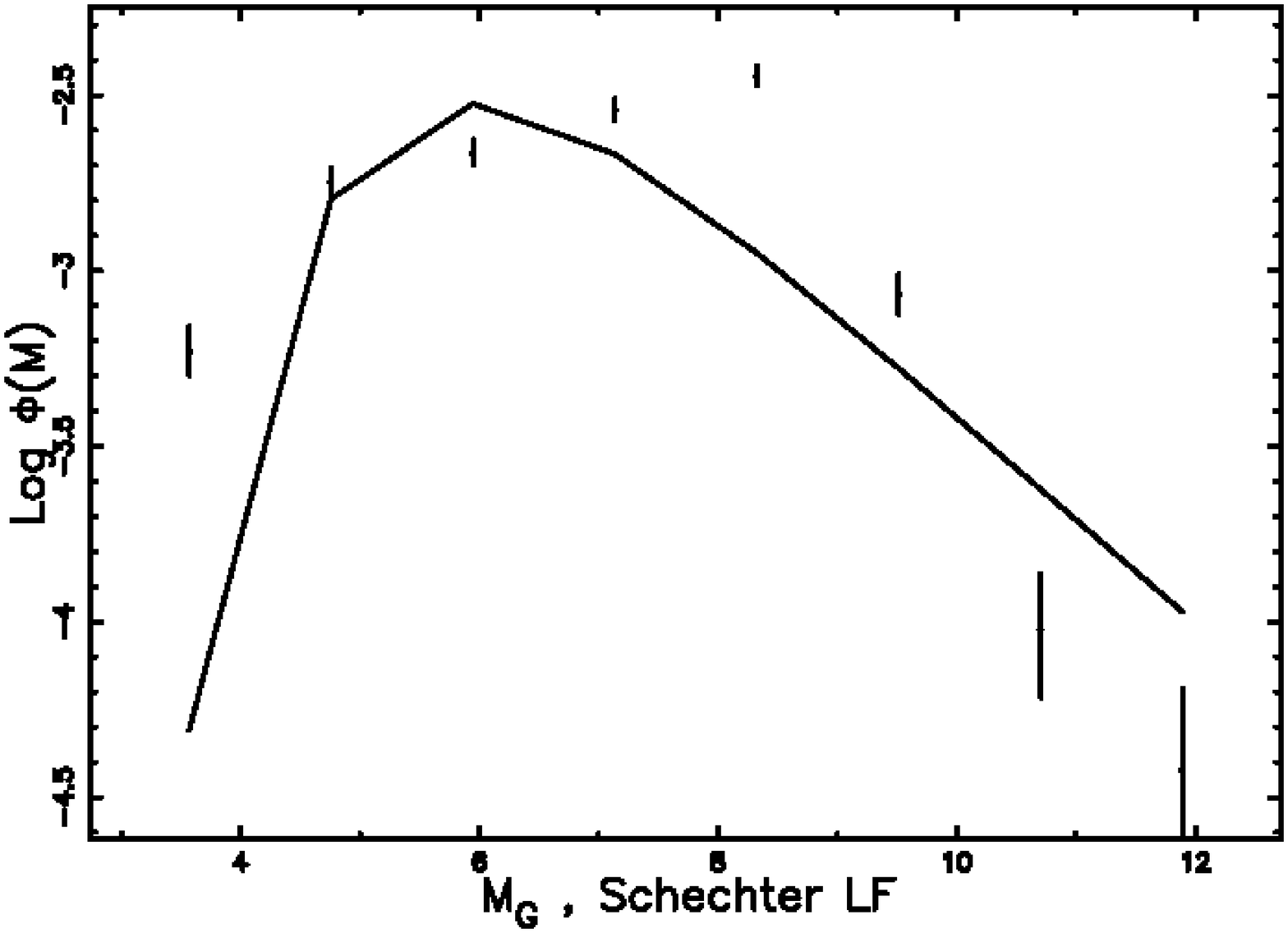}
\end{center}
\caption{
The  observed LF for stars, empty stars with error bar,
and the fit  by  the Schechter LF
when the distance covers the range   $[0\,pc , 20 \, pc ]$.
}
 \label{gaia_lf_schechter_der}%
\end{figure}

The resulting fitted curve
is displayed in Figure  \ref{gaia_lf_schechter_der}
with parameters as in
Table \ref{schechterfit}.

\begin{table}
\caption
{
Parameters  of the  Schechter LF
in the range in distance $[0\,pc , 20 \, pc ]$
when $k=3$ and $n=8$.
}
 \label{schechterfit}
 \[
 \begin{array}{ccccccc}
 \hline
 \hline
 \noalign{\smallskip}
 M^*\, (mag)
&  \Psi^* \,(pc^{-3})
&   \alpha
&   \chi^2
&  \chi_{red}^2
&  Q
&  AIC
\\
 \noalign{\smallskip}
 \hline
 5.59  & 0.0085 &  -0.26 &  166.97
&  33.39  &  3.23\,10^{-34} &  172.97
\\
 \hline
 \hline
 \end{array}
 \]
\end{table}

\subsection{The Gamma LF}

The {\em Gamma } LF, defined in $[0, \infty]$,
is
\begin {equation}
f(L;\Psi^*,L^*,c) = \Psi^*
\frac {
 \left( {\frac {L}{L^*}} \right) ^{c-1}{{\rm e}^{-{\frac {L}{L^*}}}}
}
{
L^*\Gamma  \left( c \right)
}
\label{Gammastandard}
\end {equation}
where
$\Psi^*$ is the total number of stars per pc$^3$,
\begin{equation}
\mathop{\Gamma\/}\nolimits\!\left(z\right)
=\int_{0}^{\infty}e^{{-t}}t^{{z-1}}dt
\quad ,
\end{equation}
where
    $L^*\, > \, 0$   is the scale
and $c > \, 0$    is the shape,
see  formula (17.23) in \cite{univariate1}.
The average luminosity   is
\begin{equation}
\langle f(L;\Psi^*,L^*,c) \rangle = \Psi^* L^*c  \quad.
\end{equation}
The change of parameter $(c-1)=\alpha$ allows obtaining
the same scaling as for the
Schechter LF (\ref{lf_schechter}), for more details, see
\cite{Zaninetti2016a}.
The version in absolute magnitude  is
\begin{eqnarray}
\rm
\Psi (M;\Psi^*,c,M^*)dM=
\nonumber \\
\frac
{
 0.4\,{\it \Psi^*}\, \left( {\frac {{10}^{- 0.4\,M}}{{10}^{- 0.4\,{M}^
{{\it star}}}}} \right) ^{c-1}{{\rm e}^{-{\frac {{10}^{- 0.4\,M}}{{10}
^{- 0.4\,{M}^{{\it star}}}}}}}{10}^{- 0.4\,M}\ln  \left( 10 \right)
}
{
{10}^{- 0.4\,{M}^{{\it star}}}\Gamma \left( c \right)
}
dM
\quad  .
\end{eqnarray}

\begin{figure}
\begin{center}
\includegraphics[width=7cm]{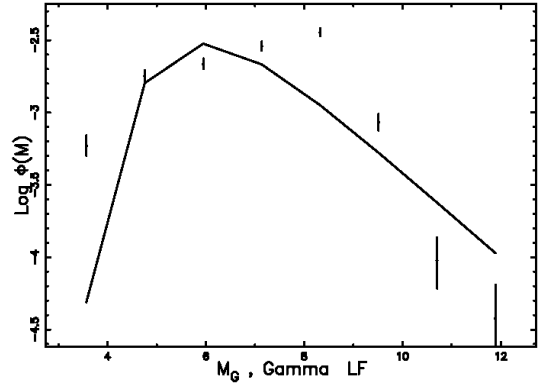}
\end{center}
\caption{
The  observed LF for stars, empty stars with error bar,
and the fit  by  the Gamma LF
when the distance covers the range   $[0\,pc , 20 \, pc ]$.
}
 \label{gaia_lf_gammac}%
\end{figure}

The resulting fitted curve
is displayed in Figure  \ref{gaia_lf_gammac}
with parameters as in
Table \ref{gammacfit}.

\begin{table}
\caption
{
Parameters  of the  Gamma LF
in the range in distance $[0\,pc , 20 \, pc ]$
when $k=3$ and $n=8$.
}
 \label{gammacfit}
 \[
 \begin{array}{ccccccc}
 \hline
 \hline
 \noalign{\smallskip}
 M^*\,  (mag)
&  \Psi^* \,(pc^{-3})
&   c
&   \chi^2
&  \chi_{red}^2
&  Q
&  AIC
\\
 \noalign{\smallskip}
 \hline
 5.59  &  0.01 &  0.73 &  166.9
&  33.39  &  3.23\,10^{-34} &  172.9
\\
 \hline
 \hline
 \end{array}
 \]
\end{table}

\section{Truncated LFs}

Here we derive the  truncated version
of the Schechter and Gamma LFs.

\label{sectiontruncated}

\subsection{The truncated Schechter LF}

The luminosity $L$ is defined
in the interval
$[L_l, L_u ]$, where the indices $l$ and $u$ mean
`lower' and `upper';
the truncated Schechter   LF, $S_T$, is
\begin {equation}
S_T(L;\Psi^*,\alpha,L^*,L_l,L_u)=
\frac
{
- \left( {\frac {L}{{\it L^*}}} \right) ^{\alpha}{{\rm e}^{-{\frac {
L}{{\it L^*}}}}}{\it \Psi^*}\,\Gamma \left( \alpha+1 \right)
}
{
{\it L^*}\, \left( \Gamma \left( \alpha+1,{\frac {L_{{u}}}{{\it
L^*}}} \right) -\Gamma \left( \alpha+1,{\frac {L_{{l}}}{{\it L^*}}
} \right)  \right)
}
\label{lf_trunc_schechter}
\quad ,
\end{equation}
where $\Gamma(a, z)$ is the incomplete Gamma function,
defined by
\begin{equation}
\mathop{\Gamma\/}\nolimits\!\left(a,z\right)
=\int_{z}^{\infty}t^{a-1}e^{-t}dt
\quad ,
\end{equation}
see  \cite{NIST2010}.
The average value  is
\begin{equation}
{ \langle S_T(L;\Psi^*,\alpha,L^*,L_l,L_u) \rangle }
=
\frac{
N
}
{
{\it L^*}  \left( \Gamma \left( \alpha+1,{\frac {L_{{u}}}{{\it
L^*}}} \right) -\Gamma \left( \alpha+1,{\frac {L_{{l}}}{{\it L^*}}
} \right)  \right)
}
\end{equation}
with
\begin{eqnarray}
N=
{\it \Psi^*}  \Bigg  ( {{\it L^*}}^{2}\Gamma \big  ( \alpha+1,{\frac
{L_{{u}}}{{\it L^*}}} \big  ) \alpha-{{\it L^*}}^{2}\Gamma \big  (
\alpha+1,{\frac {L_{{l}}}{{\it L^*}}} \big  ) \alpha+{{\it L^*}}^{
2}\Gamma \big  ( \alpha+1,{\frac {L_{{u}}}{{\it L^*}}} \big  )
\nonumber  \\
-{{
\it L^*}}^{2}\Gamma \big  ( \alpha+1,{\frac {L_{{l}}}{{\it L^*}}}
 \big  ) -{{\it L^*}}^{-\alpha+1}{{\rm e}^{-{\frac {L_{{l}}}{{\it
L^*}}}}}{L_{{l}}}^{\alpha+1}+{{\it L^*}}^{-\alpha+1}{{\rm e}^{-{
\frac {L_{{u}}}{{\it L^*}}}}}{L_{{u}}}^{\alpha+1} \Bigg  )
\times \nonumber  \\
\Gamma \big  ( \alpha+1 \big  )
\quad  .
\end{eqnarray}
The four luminosities
$L,L_l,L^*$ and $L_u$
are  connected with  the
absolute magnitudes $M$,
$M_l$, $M_u$ and $M^*$
through the following relation,
\begin{eqnarray}
\frac {L}{L_{\sun}} =
10^{0.4(M_{\sun} - M)}
\quad  ,
\frac {L_l}{L_{\sun}} =
10^{0.4(M_{\sun} - M_u)}
\quad ,
\nonumber  \\
 \frac {L^*}{L_{\sun}} =
10^{0.4(M_{\sun} - M^*)}
\,
\quad , \frac {L_u}{L_{\sun}} =
10^{0.4(M_{\sun} - M_l)}
\label{magnitudes}
\end{eqnarray}
where the indices $u$ and $l$ are inverted in
the transformation
from luminosity to absolute magnitude
and $L_{\sun}$ and  $M_{\sun}$ are  the luminosity and absolute magnitude
of the sun in the considered band.
The equivalent form  in absolute magnitude
of the truncated Schechter LF is therefore
\begin{equation}
\Psi (M;\Psi^*,\alpha,M^*,M_l,M_u)dM
=  \frac{AS}{DS}
\quad ,
\end{equation}
with
\begin{eqnarray}
AS =
- 0.4   \left( {10}^{ 0.4  {\it M^*}- 0.4  M} \right) ^{\alpha}{
{\rm e}^{-{10}^{ 0.4  {\it M^*}- 0.4  M}}} \times
\nonumber\\
{\it \Psi^*}  \Gamma
 \left( \alpha+1 \right) {10}^{ 0.4  {\it M^*}- 0.4  M} \left( \ln
 \left( 2 \right) +\ln  \left( 5 \right)  \right)
\end{eqnarray}
and
\begin{eqnarray}
DS=
\Gamma \left( \alpha+1,{10}^{- 0.4  M_{{l}}+ 0.4  {\it M^*}}
 \right) -\Gamma \left( \alpha+1,{10}^{ 0.4  {\it M^*}- 0.4  M_{{u}
}} \right)
\end{eqnarray}
The averaged absolute magnitude, $\langle M \rangle$,
is
\begin{eqnarray}
{ \langle \Psi (M;\Psi^*,\alpha,M^*,M_l,M_u) \rangle }
=   \nonumber \\
\frac{
\int_{M_l}^{M_u} M(M;\Psi^*,\alpha,L^*,L_l,L_u) M
dM
}
{
\int_{M_l}^{M_u} M(M;\Psi^*,\alpha,L^*,L_l,L_u)
dM
}
\quad  .
\label{xmtruncated}
\end{eqnarray}
More details  can be found  in  \cite{Zaninetti2017a}.

\begin{figure}
\begin{center}
\includegraphics[width=7cm]{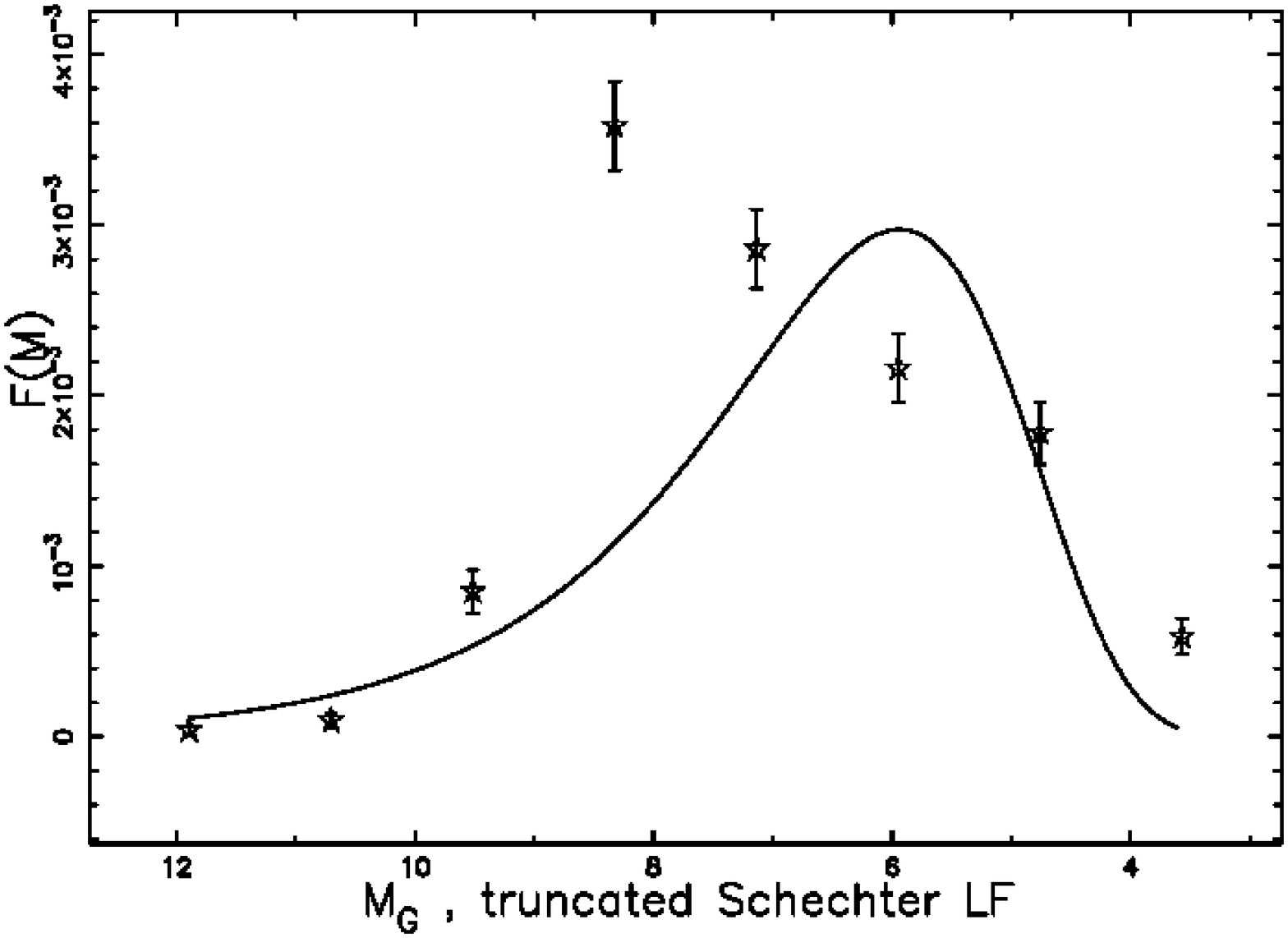}
\end{center}
\caption{
The  observed LF for stars, empty stars with error bar,
and the fit  by  the truncated Schechter LF
when the distance covers the range   $[0\,pc , 20 \, pc ]$.
}
 \label{gaia_lf_schechter_noder_trunc}%
\end{figure}

The resulting fitted curve
is displayed in Figure  \ref{gaia_lf_schechter_noder_trunc}
with parameters as in
Table \ref{truncschechterfit}.

\begin{table}
\caption
{
Parameters  of the  truncated Schechter LF
in the range in distance $[0\,pc , 20 \, pc ]$
when $k=5$ and $n=8$.
}
 \label{truncschechterfit}
 \[
 \begin{array}{ccccccccc}
 \hline
 \hline
 \noalign{\smallskip}
 M^*  
&  M_l 
&  M_u 
&  \Psi^*
&   \alpha
&   \chi^2
&  \chi_{red}^2
&  Q
&  AIC
\\
 \noalign{\smallskip}
 \hline
 5.6 &  3.56  & 11.89  & 0.0083 &  -0.26 &  167
&  55.67  &  5.54\,10^{-36} &  177
\\
 \hline
 \hline
 \end{array}
 \]
\end{table}

\subsection{The truncated Gamma LF}

The truncated Gamma LF is  defined
in the interval
$[L_l, L_u ]$
\begin {equation}
f(L;\Psi^*,L^*,c,L_l,L_u) = \Psi^*\,
k\;\left( {\frac {L}{{\it L^*}}} \right) ^{c-1}{{\rm e}^{-{\frac {L}{{
\it L^*}}}}}
\label{Gammatruncated}
\end {equation}
where
the constant $k$ is
\begin{eqnarray}
k =  \nonumber \\
\frac{c}
{
{\it L^*}\, \left(  \left( {\frac {L_{{u}}}{{\it L^*}}} \right) ^{
c}{{\rm e}^{-{\frac {L_{{u}}}{{\it L^*}}}}}-\Gamma  \left( 1+c,{
\frac {L_{{u}}}{{\it L^*}}} \right) +\Gamma  \left( 1+c,{\frac {L_{{
l}}}{{\it L^*}}} \right) - \left( {\frac {L_{{l}}}{{\it L^*}}}
 \right) ^{c}{{\rm e}^{-{\frac {L_{{l}}}{{\it L^*}}}}} \right)
}
\, .
\label{constant}
\end {eqnarray}
Its expected value   is
\begin{eqnarray}
\langle f(L;\Psi^*,L^*,c,L_l,L_u) \rangle
=
\nonumber \\
\Psi^*
\frac
{
-c \left( \Gamma  \left( 1+c,{\frac {L_{{u}}}{{\it L^*}}} \right) -
\Gamma  \left( 1+c,{\frac {L_{{l}}}{{\it L^*}}} \right)  \right) {
\it L^*}
}
{
\left( {\frac {L_{{u}}}{{\it L^*}}} \right) ^{c}{{\rm e}^{-{\frac {
L_{{u}}}{{\it L^*}}}}}-\Gamma  \left( 1+c,{\frac {L_{{u}}}{{\it
L^*}}} \right) +\Gamma  \left( 1+c,{\frac {L_{{l}}}{{\it L^*}}}
 \right) - \left( {\frac {L_{{l}}}{{\it L^*}}} \right) ^{c}{{\rm e}^
{-{\frac {L_{{l}}}{{\it L^*}}}}}
}
\quad .
\label{meanGammatruncated}
\end{eqnarray}
More details on the truncated Gamma PDF can be found
in \cite{Zaninetti2013e,Okasha2014,Zaninetti2016a}.
The  Gamma truncated LF  in magnitude is
\begin{eqnarray}
\label{lfgtmagni}
\Psi (M;\Psi^*,c,M^*,M_l,M_u) dM
=
\nonumber \\
\frac
{
 0.4\,c \left( {10}^{ 0.4\,{\it M^*}- 0.4\,M} \right) ^{c}{{\rm e}^
{-{10}^{ 0.4\,{\it M^*}- 0.4\,M}}}{\it \Psi^*}\, \left( \ln
 \left( 2 \right) +\ln  \left( 5 \right)  \right)
}
{
D
}
\end{eqnarray}
where
\begin{eqnarray}
D =  \nonumber \\
{{\rm e}^{-{10}^{- 0.4\,M_{{l}}+ 0.4\,{\it M^*}}}} \left( {10}^{-
 0.4\,M_{{l}}+ 0.4\,{\it M^*}} \right) ^{c}
\nonumber  \\
-{{\rm e}^{-{10}^{ 0.4\,
{\it M^*}- 0.4\,M_{{u}}}}} \left( {10}^{ 0.4\,{\it M^*}- 0.4\,M_
{{u}}} \right) ^{c}
\nonumber  \\
-\Gamma  \left( 1+c,{10}^{- 0.4\,M_{{l}}+ 0.4\,{
\it M^*}} \right) +\Gamma  \left( 1+c,{10}^{ 0.4\,{\it M^*}- 0.4
\,M_{{u}}} \right)
\quad  .
\end{eqnarray}
The averaged absolute magnitude, $\langle M \rangle$,
is defined numerically as in Equation \ref{xmtruncated}.

\begin{figure}
\begin{center}
\includegraphics[width=7cm]{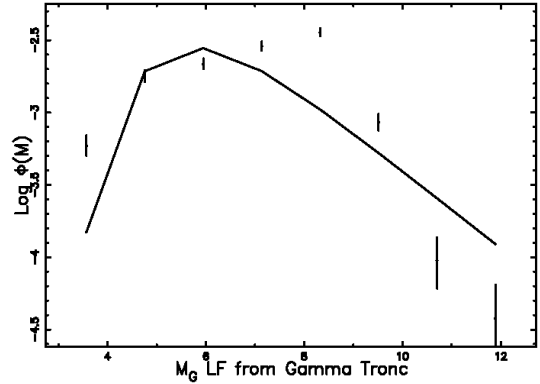}
\end{center}
\caption{
The  observed LF for stars, empty stars with error bar,
and the fit  by  the truncated Gamma LF
when the distance covers the range   $[0\,pc , 20 \, pc ]$.
}
 \label{gaia_lf_gamma_trunc}%
\end{figure}

The resulting fitted curve
is displayed in Figure  \ref{gaia_lf_gamma_trunc}
with parameters as in
Table \ref{truncGammafit}.

\begin{table}
\caption
{
Parameters  of the  truncated Gamma  LF
in the range in distance $[0\,pc , 20 \, pc ]$
when $k=5$ and $n=8$.
}
 \label{truncGammafit}
 \[
 \begin{array}{ccccccccc}
 \hline
 \hline
 \noalign{\smallskip}
 M^*   
&  M_l 
&  M_u 
&  \Psi^* 
&  c
&   \chi^2
&  \chi_{red}^2
&  Q
&  AIC
\\
 \noalign{\smallskip}
 \hline
 5.3 & 3.56 &  11.89  &  0.01 &  0.67 &  169
&  56.33 &  2.02\,10^{-36} &  179
\\
 \hline
 \hline
 \end{array}
 \]
\end{table}

\section{Distance effects}

We model  the  average  absolute
magnitude  of the stars  as  a function
of the distance,
the  photometric maximum in the number of stars for a given flux
as  a function of the distance,
and  the average distance of the stars for a given flux
in the framework of the
two truncated LFs here considered.

\label{secdistance}

\subsection{Averaged absolute magnitude}
\label{secaverage}
\begin{figure}
\begin{center}
\includegraphics[width=10cm]{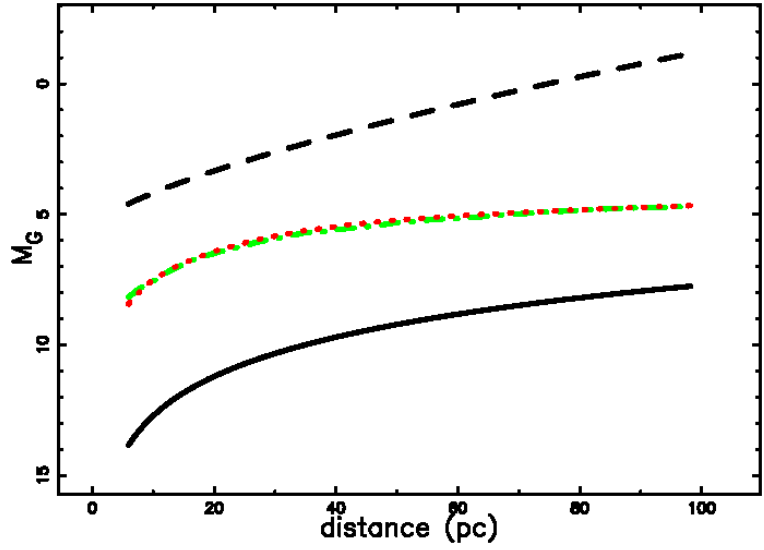}
\end{center}
\caption
{
Average observed  absolute G-magnitude
versus  distance  for Gaia (green  points),
average theoretical absolute magnitude
for truncated Schechter   LF with  $\alpha=-0.61$
as given by Equation (\ref{xmtruncated})
(dot-dash-dot red  line),
curve for the empirical
lowest absolute magnitude  at  a given
distance,
see Equation (\ref{mld}) (full black line) and
the theoretical  curve
for the highest absolute magnitude  at  a given
distance (dashed black line),
see  Equation (\ref{mabsgupper}).
}
\label{gaia_xmd_schechter_trunc}
\end{figure}

In order to model the influence
of the distance $d$ in pc on the LF, an empirical
variable  lower  bound in absolute magnitude,
$M_l$,
has been introduced,
\begin{equation}
M_l(d)  =5.53  -0.27 \, d ^{0.7}
\label{mld}
\quad .
\end{equation}
The upper bound, $M_u$  was already fixed
by the nonlinear equation (\ref{mabsgupper}).
A second distance  correction  is
\begin{equation}
M^*= M_u(d) - 2.8  - 5.2 \exp{- \frac{d}{100} }
\quad ,
\label{mstarcorrection}
\end{equation}
where
$M_u(d)$  has  been defined in
Equation (\ref{mabsgupper}).
Figure \ref{gaia_xmd_schechter_trunc}
compares
the theoretical
average absolute magnitudes for the truncated  Schechter LF
with the observed ones;
the value of $M^*$ in Equation (\ref{mstarcorrection})
minimizes the difference between the two curves.

\begin{figure}
\begin{center}
\includegraphics[width=10cm]{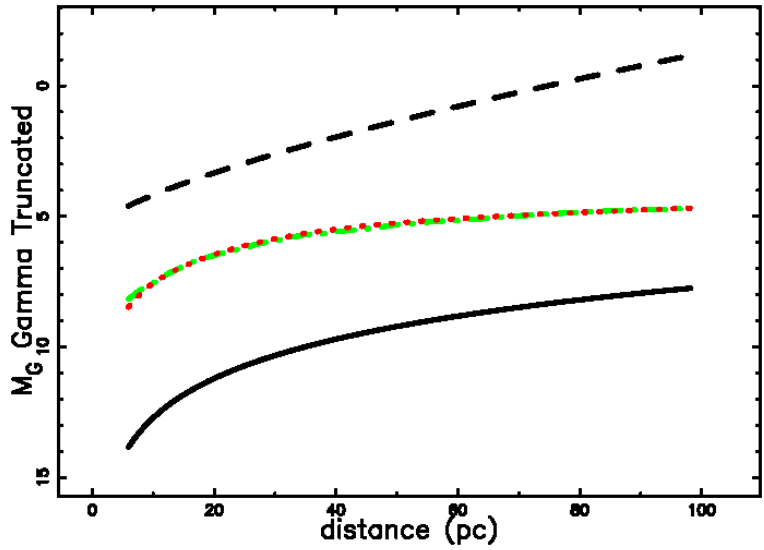}
\end{center}
\caption
{
Average observed  absolute G-magnitude
versus  distance  for Gaia (green  points),
average theoretical absolute magnitude
for truncated Gamma   LF with  $c=0.38$
as given by the analogue of equation (\ref{xmtruncated})
(dot-dash-dot red  line),
theoretical curve for the empirical
lowest absolute magnitude  at  a given
distance,
see Equation (\ref{mld}) (full black line) and
the theoretical  curve
for the highest absolute magnitude  at  a given
distance (dashed black line),
see  Equation (\ref{mabsgupper}).
}
\label{gaia_xmd_gamma_trunc}
\end{figure}

Conversely Figure \ref{gaia_xmd_gamma_trunc}
compares
the theoretical
average absolute magnitudes for the truncated  Gamma LF
with the observed ones;
also here
the value of $M^*$ obtained from
Equation (\ref{mstarcorrection})
minimizes the difference between the two curves.

\subsection{The photometric maximum}

The definition of the flux, $f$,
is
\begin{equation}
f  = \frac{L}{4 \pi r^2}
\quad ,
\label{flux}
\end{equation}
where $r$ is the distance and $L$ the luminosity of the star.
The joint distribution in distance, {\it r},
and  flux, {\it f},  for the number of stars is
\begin{equation}
\frac{dN}{d\Omega dr df} = \frac{1}{4 \,\pi}
\int_0^{\infty } 4 \pi r^2  dr
\Phi(\frac{L}{L^*}) \delta (f-\frac{L}{4\,\pi\,r^2} )
\quad ,
\label{nldef}
\end{equation}
were the factor  ($\frac{1}{4 \pi}$)
converts the number density
into density for solid angle
and the Dirac delta function  selects  the required flux.
We now apply the sifting properties of the delta function,
see \cite{Bracewell2000},
to the case of the  Schechter  LF
as  given by formula  \ref{lf_schechter}
\begin{equation}
\frac{dN}{d\Omega dr df} =
\frac{1}{L^*}\,
4\,\pi\,{r}^{4}{\it \Phi^*}\, \left( 4\,{\frac {\pi\,f{r}^{2}}{{\it L^*}}
} \right) ^{\alpha}{{\rm e}^{-4\,{\frac {\pi\,f{r}^{2}}{{\it L^*}}}}
}
\quad .
\label{nfunctionrschechter}
\end{equation}

We now introduce the critical radius $r_{crit}$
\begin{equation}
r_{crit}=
\frac{1}{2}
\,{\frac {\sqrt {{\it L^* }}}{\sqrt {\pi}\sqrt {f}}}
\quad  .
\end{equation}
Therefore
the joint distribution in distance
and  flux becomes
\begin{equation}
\frac{dN}{d\Omega dr df} =
\frac{1}{L^*} \,
4\,\pi\,{r}^{4}{\it \Phi^*}\, \left( {\frac {{r}^{2}}{{{\it r_{crit}}}^{2}}}
 \right) ^{\alpha}{{\rm e}^{-{\frac {{r}^{2}}{{{\it r_{crit}}}^{2}}}}}
\quad .
\label{nfunctionrschechter_rcrit}
\end{equation}
The above  number of stars
has a maximum at $r=r_{max}$:
\begin{equation}
r_{max}= \sqrt {2+\alpha}{\it r_{crit}}
\quad  ,
\label{rmaxrcrit}
\end{equation}
and the average  distance  of the stars,
$ { \langle r \rangle} $,
is
\begin{equation}
\langle r \rangle
={\frac {{\it r_{crit}}\,\Gamma \left( 3+\alpha \right) }
{\Gamma \left( \frac{5}{2} +\alpha \right) }}
\quad .
\label{raverageflux}
\end{equation}

\begin{figure}
\begin{center}
\includegraphics[width=6cm]{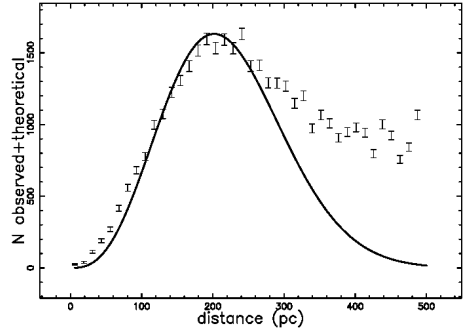}
\end {center}
\caption
{
The stars of Gaia with
$ 1243361.38
 \,(e-/s)   \leq
f \leq 1450291.3 \, (e-/s)  $
or  
$ 10.121 \, (mag)  \leq G \,(mag) \, \leq 10.288$ 
are  organized by frequency versus
distance,  (empty circles);
the error bar is given by the square root of the frequency.
The maximum frequency of the observed stars is
at  $d=247$\ pc.
The full line is the theoretical curve
generated by
$\frac{dN}{d\Omega dr df}$
as given by the application of the Schechter LF
which  is Equation (\ref{nfunctionrschechter})
and the theoretical  maximum is at
$d=274$\ pc.
The parameters are
$L^*= 5 \,10^{11}$\ (e-/s)$\times$ pc$^2$   and
$\alpha$ =-0.62.
Case of the  Schechter  LF.
}
          \label{gaia_max_uno}%
    \end{figure}

Figure \ref{gaia_max_uno}
presents the number of  stars  observed in Gaia
as a function  of the distance  for  a given
window in the flux, as well as the theoretical curve.

\begin{figure}
\begin{center}
\includegraphics[width=6cm]{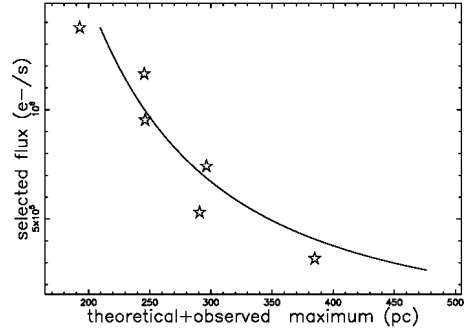}
\end {center}
\caption
{
Position of the observed maximum as function
of the flux  (empty stars) and theoretical
average value  as  given by Equation
(\ref{raverageflux})
for the Schechter LF (full  line).
The parameters are
$L^*= 5.5 \,10^{11}$\ (e-/s)$\times$ pc$^2$   and
$\alpha$ =-0.62.
Case of the  Schechter  LF.
}
          \label{gaia_max_molti}%
    \end{figure}

Figure \ref{gaia_max_molti}  presents
the  observed position of the
maximum of the number of stars
as  a function of the flux.

\begin{figure}
\begin{center}
\includegraphics[width=6cm]{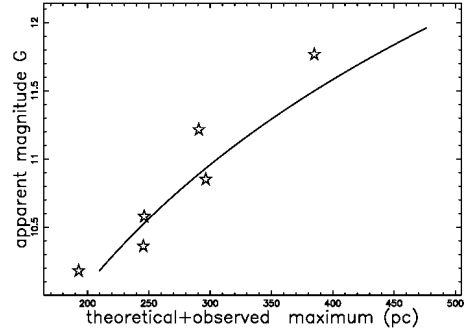}
\end {center}
\caption
{
Position of the observed maximum as function
of the apparent magnitude, G,  (empty stars) and theoretical
average value  as  given by Equation
(\ref{raverageflux})
for the Schechter LF (full  line).
The parameters are the same of Figure 
\ref{gaia_max_molti}.
Case of the  Schechter  LF.
}
          \label{molti_max_g}%
    \end{figure}

In order to shift to more familiar 
variables Figure \ref{molti_max_g} reports
the position of the above maximum as 
function of the apparent Gaia magnitude

\begin{figure}
\begin{center}
\includegraphics[width=6cm]{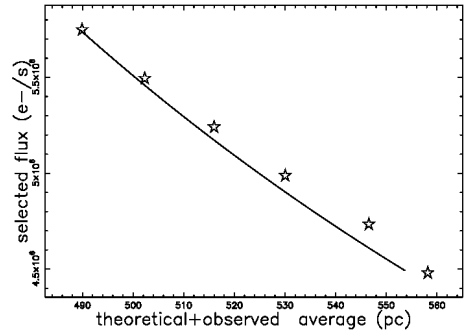}
\end {center}
\caption
{
Position of the average distance of the stars
as function
of the flux  (empty stars) and theoretical
curve  as  given by Equation  (\ref{rmaxrcrit})
(full  line) for the Schechter LF.
The parameters are
$L^*= 1.3 \,10^{13}$\ (e-/s)$\times$\ pc$^2$   and
$\alpha$ =-0.62.
Case of the  Schechter  LF.
}
          \label{gaia_ave_schechter}%
    \end{figure}

\begin{figure}
\begin{center}
\includegraphics[width=6cm]{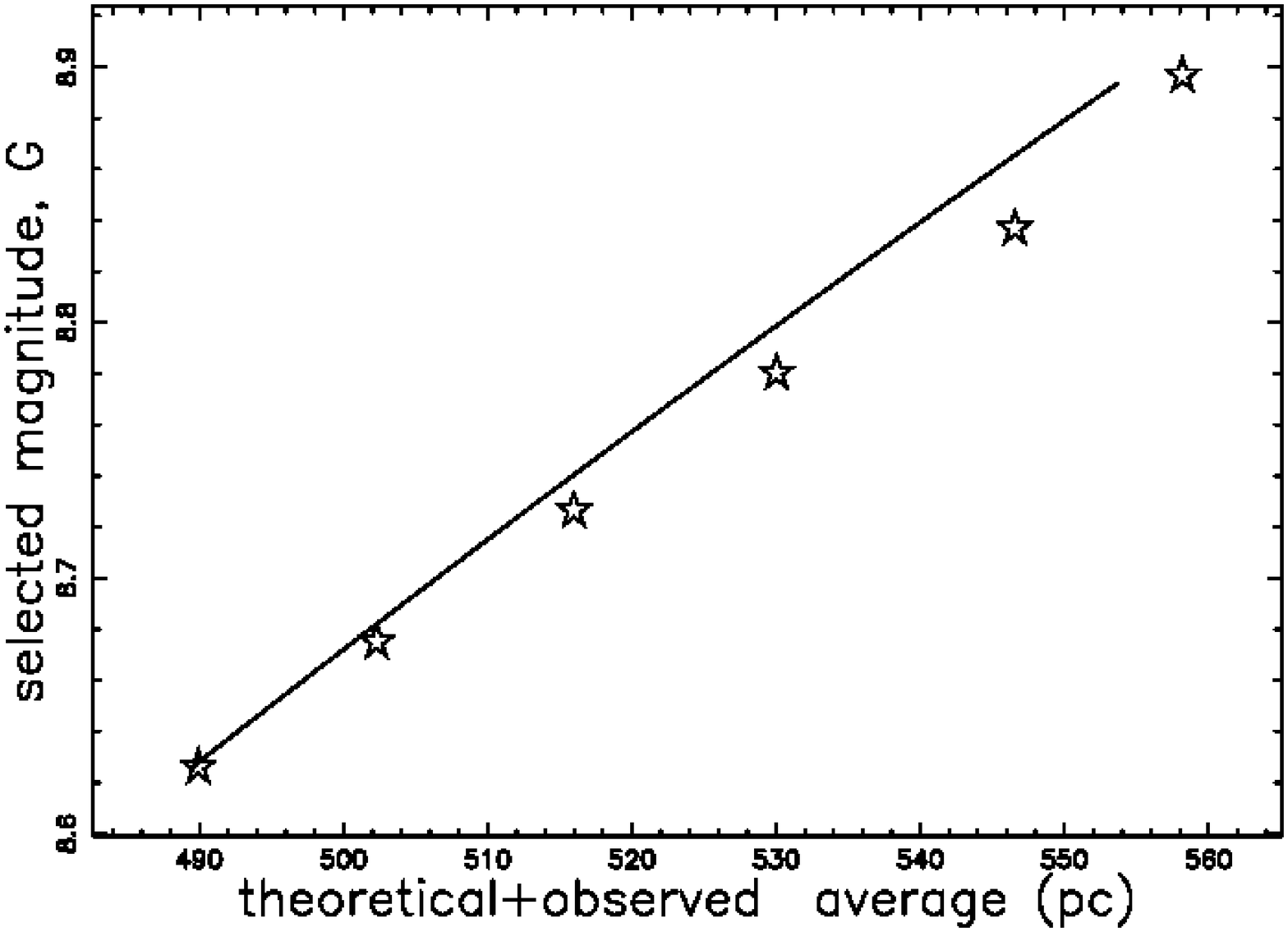}
\end {center}
\caption
{
Position of the average distance of the stars
as function
of the apparent magnitude, G,   (empty stars) and theoretical
curve  as  given by Equation  (\ref{rmaxrcrit})
(full  line) for the Schechter LF.
The parameters are the same of Figure \ref{gaia_ave_schechter}.
Case of the  Schechter  LF.
}
          \label{ave_schechter_g}%
    \end{figure}

Figures \ref{gaia_ave_schechter}  
and \ref{ave_schechter_g} 
present
the  observed  average value
of the number of stars
as  a function of the flux and apparent magnitude.

In the case of the Gamma LF,
the  maximum  in the number of stars is at
\begin{equation}
r_{max}= \sqrt {c+1}{ r_{crit}}
\quad  ,
\label{rmaxrcritGamma}
\end{equation}
and the average  distance  of the stars
$ { \langle r \rangle} $,
is
\begin{equation}
\langle r \rangle
={\frac {{ r_{crit}}\,c \left( c+1 \right) \Gamma \left( c \right) }{
\Gamma \left( \frac{3}{2}+c \right) }}
\quad .
\label{raveragefluxGamma}
\end{equation}

\section{Conclusions}

{\it Standard LFs}.  
The Schechter function and the Gamma PDF
can model  the LF for stars,
see Tables \ref{schechterfit}
and \ref{gammacfit}
as well as Figures \ref{gaia_lf_schechter_der}
and \ref{gaia_lf_gammac},
but
the values of the involved parameters depend
on the chosen distance.
\newline
{\it Truncated LFs.} The truncated  Schechter function
and the truncated Gamma LF can model
the averaged absolute magnitude
as a function of the distance,
see Figures \ref{gaia_xmd_schechter_trunc}
and         \ref{gaia_xmd_gamma_trunc}.
As an example, four analytical equations have been used
in the case of the truncated  Schechter LF:
(i) the average theoretical absolute magnitude
for the truncated Schechter LF,
see Equation (\ref{xmtruncated}),
(ii) an empirical expression for the
lowest absolute magnitude  at  a given
distance, see Equation (\ref{mld}),
(iii) a theoretical  curve
for the highest absolute magnitude  at  a given
distance,
see  Equation (\ref{mabsgupper}),
and (iv) a   distance dependence
expression  for  $M^*$
as  given by Equation \ref{mstarcorrection}.
The above four equations model the Malmquist bias.
\newline
{\it Photometric maximum}
The number of stars as a function of the distance
presents a maximum which is a function
of the flux, see Figures
\ref{gaia_max_uno} and \ref{gaia_max_molti}
for the Schechter LF.
The theoretical and observed
average distance  of the stars  are also functions
of the selected flux, see Figure \ref{gaia_ave_schechter}.
\newline
{\it Topics not covered}
The treatment  here adopted deals
with an homogeneous distribution of stars  and therefore the
the vertical scale-heights
are not covered, see \cite{Just2015}.

\section*{Acknowledgments}

This work has made use of data from the European Space Agency (ESA)
mission {\it Gaia}
\\
(\url{https://www.cosmos.esa.int/gaia}),
processed by
the {\it Gaia} Data Processing and Analysis Consortium (DPAC,
\url{https://www.cosmos.esa.int/web/gaia/dpac/consortium}). Funding
for the DPAC has been provided by national institutions, in particular
the institutions participating in the {\it Gaia} Multilateral Agreement.
\\
\section*{Bibliography}

\begin{thebibliography}{10}
\expandafter\ifx\csname url\endcsname\relax
  \def\url#1{{\tt #1}}\fi
\expandafter\ifx\csname urlprefix\endcsname\relax\def\urlprefix{URL }\fi
\providecommand{\eprint}[2][]{\url{#2}}

\bibitem{Malmquist_1920}
{Malmquist } K~G 1920 {A study of the stars of spectral type A } {\em Lund
  Medd. Ser.~II\/} {\bf 22}, 1

\bibitem{Malmquist_1922}
{Malmquist } K~G 1922 {On some relations in stellar statistics } {\em Lund
  Medd. Ser.~I\/} {\bf 100}, 1

\bibitem{Malmquist_1936}
{Malmquist} K~G 1936 {Investigations on the stars in high galactic latitudes
  II. Photographic magnitudes and colour indices of about 4500 stars near the
  north galactic pole.} {\em Stockholms Observatoriums Annaler\/} {\bf 12}, 7.1

\bibitem{Binney1998}
{Binney} J and {Merrifield} M 1998 {\em {Galactic astronomy}\/} (Princeton, NJ:
  Princeton University Press)

\bibitem{Butkevich2005}
{Butkevich} A~G, {Berdyugin} A~V and {Teerikorpi} P 2005 {Statistical biases in
  stellar astronomy: the Malmquist bias revisited} {\em \mnras\/} {\bf 362},
  321

\bibitem{Eddington1914}
{Eddington} A~S 1914 {\em {Stellar movements and the structure of the
  universe}\/} (London: Macmillan and co.)

\bibitem{Jaschek1985}
{Jaschek} C and {Gomez} A~E 1985 {The Malmquist correction} {\em \aap\/} {\bf
  146}, 387

\bibitem{Wielen1974}
{Wielen} R 1974 {The kinematics and ages of stars in Gliese's catalogue} {\em
  Highlights of Astronomy\/} {\bf 3}, 395

\bibitem{Flynn2006}
{Flynn} C, {Holmberg} J, {Portinari} L, {Fuchs} B and {Jahrei{\ss}} H 2006 {On
  the mass-to-light ratio of the local Galactic disc and the optical luminosity
  of the Galaxy} {\em \mnras\/} {\bf 372}, 1149 (\textit{Preprint}
  \eprint{astro-ph/0608193})

\bibitem{Just2015}
{Just} A, {Fuchs} B, {Jahrei{\ss}} H, {Flynn} C, {Dettbarn} C and {Rybizki} J
  2015 {The local stellar luminosity function and mass-to-light ratio in the
  near-infrared} {\em \mnras\/} {\bf 451}, 149 (\textit{Preprint}
  \eprint{1504.05808})

\bibitem{GAIA2016a}
{Gaia Collaboration}, {Prusti} T, {de Bruijne} J~H~J, {Brown} A~G~A,
  {Vallenari} A, {Babusiaux} C, {Bailer-Jones} C~A~L, {Bastian} U, {Biermann}
  M, {Evans} D~W and et~al 2016 {The Gaia mission} {\em \aap\/} {\bf 595} A1
  (\textit{Preprint} \eprint{1609.04153})

\bibitem{GAIA2016b}
{Gaia Collaboration}, {Brown} A~G~A, {Vallenari} A, {Prusti} T, {de Bruijne}
  J~H~J, {Mignard} F, {Drimmel} R, {Babusiaux} C, {Bailer-Jones} C~A~L,
  {Bastian} U and et~al 2016 {Gaia Data Release 1. Summary of the astrometric,
  photometric, and survey properties} {\em \aap\/} {\bf 595} A2
  (\textit{Preprint} \eprint{1609.04172})

\bibitem{Stassun2016}
{Stassun} K~G and {Torres} G 2016 {Evidence for a Systematic Offset of -0.25
  mas in the Gaia DR1 Parallaxes} {\em \apjl\/} {\bf 831} L6

\bibitem{GAIA2017b}
{Evans} D~W, {Riello} M, {De Angeli} F, {Busso} G, {van Leeuwen} F, {Jordi} C,
  {Fabricius} C, {Brown} A~G~A, {Carrasco} J~M, {Voss} H and et~al 2017 {Gaia
  Data Release 1. Validation of the photometry} {\em \aap\/} {\bf 600} A51
  (\textit{Preprint} \eprint{1701.05873})

\bibitem{GAIA2016c}
{Carrasco} J~M, {Evans} D~W, {Montegriffo} P, {Jordi} C, {van Leeuwen} F,
  {Riello} M, {Voss} H, {De Angeli} F, {Busso} G and et~al 2016 {Gaia Data
  Release 1. Principles of the photometric calibration of the G band} {\em
  \aap\/} {\bf 595} A7 (\textit{Preprint} \eprint{1611.02036})

\bibitem{GAIA2017a}
{van Leeuwen} F, {Evans} D~W, {De Angeli} F, {Jordi} C, {Busso} G, {Cacciari}
  C, {Riello} M, {Pancino} E, {Altavilla} G and et~al 2017 {Gaia Data Release
  1. The photometric data} {\em \aap\/} {\bf 599} A32 (\textit{Preprint}
  \eprint{1612.02952})

\bibitem{press}
{Press} W~H, {Teukolsky} S~A, {Vetterling} W~T and {Flannery} B~P 1992 {\em
  {Numerical Recipes in FORTRAN. The Art of Scientific Computing}\/}
  (Cambridge, UK: Cambridge University Press)

\bibitem{Akaike1974}
{Akaike} H 1974 {A new look at the statistical model identification} {\em IEEE
  Transactions on Automatic Control\/} {\bf 19}, 716

\bibitem{Liddle2004}
{Liddle} A~R 2004 {How many cosmological parameters?} {\em \mnras\/} {\bf 351},
  L49

\bibitem{Godlowski2005}
{Godlowski} W and {Szydowski} M 2005 {Constraints on Dark Energy Models from
  Supernovae} in M~{Turatto}, S~{Benetti}, L~{Zampieri} and W~{Shea}, eds, {\em
  1604-2004: Supernovae as Cosmological Lighthouses\/} vol 342 of {\em
  Astronomical Society of the Pacific Conference Series\/} pp 508--516

\bibitem{schechter}
{Schechter} P 1976 {An analytic expression for the luminosity function for
  galaxies.} {\em \apj\/} {\bf 203}, 297

\bibitem{univariate1}
{Johnson} N~L, {Kotz} S and {Balakrishnan} N 1994 {\em {Continuous univariate
  distributions. Vol. 1. 2nd ed.}\/} (New York: {Wiley })

\bibitem{Zaninetti2016a}
{Zaninetti} L 2016 Pade approximant and minimax rational approximation in
  standard cosmology {\em Galaxies\/} {\bf 4}(1), 4 ISSN 2075-4434
  \urlprefix\url{http://www.mdpi.com/2075-4434/4/1/4}

\bibitem{NIST2010}
Olver F~W~J~e, Lozier D~W~e, Boisvert R~F~e and Clark C~W~e 2010 {\em {NIST
  handbook of mathematical functions.}\/} (Cambridge: {Cambridge University
  Press. })

\bibitem{Zaninetti2017a}
{Zaninetti} L 2017 A left and right truncated schechter luminosity function for
  quasars {\em Galaxies\/} {\bf 5}(2), 25

\bibitem{Zaninetti2013e}
{Zaninetti} L 2013 {A right and left truncated gamma distribution with
  application to the stars } {\em Advanced Studies in Theoretical Physics\/}
  {\bf 23}, 1139

\bibitem{Okasha2014}
{Okasha} M~K and {Alqanoo} I~M 2014 {Inference on The Doubly Truncated Gamma
  Distribution For Lifetime Data} {\em International Journal Of Mathematics And
  Statistics Invention\/} {\bf 2}, 1

\bibitem{Bracewell2000}
{Bracewell} R~N 2000 {\em {The Fourier transform and its applications}\/} (New
  York: McGraw-Hill)

\end{thebibliography}
\providecommand{\newblock}{}

\providecommand{\newblock}{}
\end{document}